\documentclass[12pt,a4paper]{article}
\usepackage{amsmath}
\usepackage{amssymb}
\usepackage[dvips]{graphicx}
\usepackage{rotating}

\def\Tr{\mathop{\rm Tr}\nolimits}
 
\title{
The Bogomolny Equations and Solutions for
Einstein-Yang-Mills-Dilaton-$\sigma$ Models}

\author{H. W. Braden\thanks{E-mail:hwb@maths.ed.ac.uk}\ \ and V. Varela
\thanks{E-mail:varela@maths.ed.ac.uk}\\
\normalsize
\em Department of Mathematics and Statistics,\\
\normalsize
\em The University of Edinburgh, \\
\normalsize
\em King's Buildings, Mayfield Road,\\
\normalsize
\em Edinburgh EH9 3JZ, UK \\
}
 
\begin{document}
 
\renewcommand{\thepage}{}
\begin{titlepage}
 
\maketitle
\vskip-9.5cm
\hskip10.4cm
MS-98-006
\vskip.2cm
\hskip10.4cm
\sf hep-th/9804204 \rm
\vskip8.8cm
\begin{abstract}
We derive Bogomolny equations for an 
Einstein-Yang-Mills-dilaton-$\sigma$ model (EYMD-$\sigma$) on a 
static spacetime, showing that the Einstein equations  are satisfied
if and only if the associated (conformally scaled) three-metric is flat. 
These are precisely the  static metrics for which
which super-covariantly constant spinors exist.  We study some  general
properties of these equations and then  consider the problem of obtaining 
axially symmetric solutions for the gauge group $SU(2)$.
\end{abstract}
\vfill
\end{titlepage}
\renewcommand{\thepage}{\arabic{page}}
 
\section{Introduction}
The Bogomolny equations \cite{bogo} have played a central role in the
search for solutions of the Yang-Mills-Higgs (YMH) system of equations
in the limit of vanishing self-interactions of the Higgs field. Because of
their lower order, the Bogomolny differential equations prove a major
simplification to the problem of finding analytical solutions for the
YMH equations. It is natural therefore to seek similar
equations in the study of Einstein-Yang-Mills-Higgs and related
systems in which the coupling to gravity increases the (already high) 
non-linearities of the coupled, second order equations. This paper presents 
several new results along such lines.

The investigation of curved-space Bogomolny equations has interested many
people. A modified version of the Euclidean Bogomolny equations was considered 
by Comtet \cite{comtet} very early on while studying Yang-Mills-Higgs (YMH) 
systems 
(in the Prasad-Sommerfield limit) on fixed, static, curved-space backgrounds
utilising a spherically symmetric ansatz. Here a system of first order
Bogomolny type equations was proposed that implied the YMH equations
of motion provided the metric satisfied a differential constraint.
After this, in an important work, Comtet, Forg\'{a}cs and Horv\'{a}thy
\cite{comforhor} treated the general fixed, static, curved-space background
with no (spatial) symmetry assumptions. Again first order
Bogomolny type equations were found that yielded the YMH equations
of motion provided the metric satisfied the differential constraint:
\begin{equation}
\Delta \ln \sqrt{|g_{00}|}=0.
\label{diffconstraint}
\end{equation}
Going beyond Comtet's earlier work, these authors now asked about the 
compatibility of this constraint with the Einstein equations (here
only considering the spherically symmetric case). Compatibility was found
possible only for a very special physical solution: the Higgs field was
required to take its vacuum value $\Tr \phi\sp2=1$ and the metric described
an extreme Reissner-Nordstr\"{o}m black hole.
As we shall see, this first constraint (that $\Tr \phi\sp2=1$, which may
be viewed as a sigma model constraint) is  fundamental, and appears either
implicitly or explicitly in most authors work.

Balakrishna and Wali \cite{bawa}, working 
with an Einstein-Yang-Mills-Higgs (EYMH) system with non-minimal $R\Tr \phi\sp2$
coupling, and assuming a conformstatic metric with flat three-geometry,
also obtained the same Bogomolny equations of Comtet {\it et al.}
Here the Einstein equations were found to be satisfied as a consequence
of the Bogomolny equations provided $\Tr \phi\sp2=1$ and the choice of units
$4\pi G=1$ was made. The sigma model constraint in fact means that there is no
difference between minimal and non-minimal gravitational coupling for the
solutions being considered; the choice of units corresponds to a balance
between the attractive gravitational and repulsive gauge force of like
particles, and will be commented upon further in the sequel.
Balakrishna and Wali also examined the large and small $r$ asymptotics
of the resulting metric, assuming spherical symmetry.
In a similar vein Cho {\it et al.}\cite{cho} considered a (now minimally
coupled) Einstein-Yang-Mills-Higgs-Dilaton (EYMHD) system for a static
metric with flat three-geometry, their work assuming the  constant
$\Tr \phi\sp2$ sigma model constraint. Analogous Bogomolny equations  to those
found earlier (but incorporating the dilaton) were obtained. Various 
multipole solutions generalising the Majumdar-Papapetrou \cite{maj, pap}
electrovac solutions were found, as well as the Gross-Perry-Sorkin 
multi-monopole solution obtained from a five dimensional Kaluza-Klein theory
\cite{GP, DS, sork}.

Most recently Forg\'{a}cs {\it et al.} \cite{forhorhor} have made explicit
the recurring sigma model constraint noted above by considering the
coupling of an Einstein-Yang-Mills system to a gauged sigma model 
(EYM-$\sigma$) or an Einstein-Yang-Mills-dilaton system with
analogous coupling (EYMD-$\sigma$).
They view this as the infinite mass limit of an EYMH system: the infinite
mass forcing the length of the Higgs to assume its vacuum minimum.
The key point is that when coupling to a sigma model field $n$, the
differential constraint (\ref{diffconstraint}) is replaced by
\begin{equation}
\Delta \ln \sqrt{|g_{00}|}= \Tr (D_in D\sp{i}n ),
\label{newconstraint}
\end{equation}
and the Yang-Mills and sigma model equations of motion follow from a 
Bogomolny equation if and only if
the metric satisfies this equation. Now (again with units  $4\pi G=1$)  
they show that (\ref{newconstraint}) is equivalent to the 
(00) component of the Einstein equations for a general static metric.
Although these authors do not prove the compatibility of their 
Bogomolny equation  with the remainder of the Einstein equations they do
present a particular case for which this compatibility holds. This
particular case corresponds to spherical symmetry and the t'Hooft ansatz:
the resulting ODE's corresponds with those found in \cite{bawa} noted
above, which shows that the issue of the compatibility is nontrivial.
Most recently Viet and Wali \cite{viwa} adopting a spherically
symmetric ansatz show the compatibility of the
Bogomolny equation with the equations of motion and Einstein equations
of a non-minimally coupled EYMH system. Here again a sigma model
constraint is imposed, and we again encounter the same equations of 
Forg\'{a}cs {\it et al.} but now the full compatibility with the
Einstein equations is shown within their ansatz.

In this paper we follow Forg\'{a}cs {\it et al.} by initially
considering an EYM-$\sigma$ model on an arbitrary, static spacetime.
Our first result is to show that, with the assumption of static fields and 
units so that $4\pi G=1$, the Bogomolny equations  
(see (\ref{Bogomolny1}) or (\ref{Bogomolny2})) yield  all of
the field equations if and only if the (appropriately conformally scaled)
three-metric is flat. 
This at once answers the question as to the
compatibility of the Bogomolny equations with the remaining Einstein
equations  and justifies the ansatz of the various authors mentioned
above. Interestingly the space-times we find Bogomolny equations for
are precisely those  static space-times for which super-covariantly
constant spinors exist \cite{gibbonshull}. This is consistent with the
arguments of \cite{HlSpect}  who associate supersymmetric extensions
to theories exhibiting Bogomolny bounds and suggests a supersymmetric extension
exists to this theory.
Having obtained our Bogomolny equations we proceed  to  investigate these
in section three. First we show that an auxiliary  magnetostatic problem
may be associated with the equations for any gauge group. This is
particularly relevant for situations where the sigma model field is 
covariantly constant and leads to the Majumdar-Papapetrou electrovac
solutions noted above. To examine the case of non-covariantly constant 
solutions we focus on the case of axially symmetric  charge one
solutions for the gauge group $SU(2)$; the ansatz we
adopt readily incorporates the spherically symmetric ansatz of the
earlier works and we recover these.
Here we also give various improved asymptotic solutions
and are able to show that an asymptotically flat, spherically
symmetric black hole solution with regular gauge fields satisfying the
Bogomolny equations is necessarily
an extreme Reissner-Nordstr\"{o}m black hole.
Section four extends our earlier analysis to include a dilaton
coupling: up to some physical redefinitions we find that the Bogomolny
equations (and consequently their solutions) are unaltered.
Our penultimate section considers axially symmetric $SU(2)$ solutions
to the Bogomolny equations with charge (or winding number) greater than one.
Here we can illustrate covariantly constant solutions with arbitrary charge 
but are unable to find solutions to the ansatz of various authors for the
non-covariantly constant case.
We conclude with a brief discussion of our results.
An appendix comments on the singularities of the solutions under 
consideration.

\section{The Bogomolny Equations}
Our study of the EYM-$\sigma$ system is based upon the action 
\begin{equation}
S=\int d^{4}x\sqrt{|g|}
\bigg[ \frac{1}{16\pi G}R
 -\frac{1}{4} \Tr F_{\mu\nu}F^{\mu\nu}
 -\frac{1}{2} \Tr  D_{\mu}n D^{\mu} n
+ \frac{\lambda}{2}\left(\Tr n\sp2-1\right)
\bigg].
\label{action}
\end{equation} 
Here $G$ is the gravitational constant, $R$ the Ricci scalar
associated to the spacetime metric $g_{\mu\nu}$ 
and the scalar field
$n$ is in the adjoint representation of the gauge group with
associated field strength $F$. Indices $\mu$, $\nu$, ... run from 0 to
3 and we are working with a signature $(-+++)$.
The Lagrange multiplier $\lambda$ of the final term in the action
imposes the $\sigma$-model constraint. This action has been considered
previously in Ref.\cite{forhorhor} as the infinite mass limit of
spontaneously broken gauge theories with adjoint Higgs fields.

The field equations derived from (\ref{action}) are
\begin{equation}
\frac{1}{\sqrt{|g|}}D_{\mu}\left(\sqrt{|g|}F^{\mu\nu}\right)
=\left[n,D^{\nu}n\right],
\label{ym}
\end{equation} 
\begin{equation}
\frac{1}{\sqrt{|g|}}D_{\mu}\left(\sqrt{|g|}D^{\mu}n\right)=
-\bigg(\Tr D^{\mu}nD_{\mu}n\bigg)\,n,
\label{sf}
\end{equation}
\begin{equation}
R_{\mu\nu}=8\pi G\left(T_{\mu\nu}-\frac{1}{2}g_{\mu\nu}T\right),
\label{eins}
\end{equation} 
where $R_{\mu\nu}$ is the Ricci tensor, $T_{\mu\nu}$ is the total
energy-momentum tensor associated to the gauge field
and the $\sigma$-model, and $T=T^{\mu}{}_{\mu}$. (See
\cite{forhorhor} for details.)

We restrict our attention to the static, purely magnetic case
($A_{0}=0$) of this theory and for the purpose of this paper
assume a static metric parameterised as
\begin{equation}
ds\sp2 = g_{00}\, dt\sp2 +g_{ij}dx\sp{i} dx\sp{j}=
-V\sp2 dt\sp2 +\frac{h_{ij}}{V\sp2}dx\sp{i} dx\sp{j}.
\label{metric}
\end{equation}
Comment on the extension to a stationary metric will be made in the
sequel and here $i,j$ run from $1$ to $3$.
For such a metric $R_{0i}=0$, and we note the identity \cite{synge}[p 339]
\begin{equation}
R_{00}=-V \square_{g}V=-V\sp3 \square_{h}V+V\sp2 h\sp{ij}\partial_i V\,
\partial_j V,
\label{ric00}
\end{equation}
where $\square_{g(h)}$ is the scalar Laplacian with respect to the three-metric
$g_{ij}$ (or $h_{ij}$).
Since the time derivative of
every field is assumed to be zero we may construct a reduced action $\cal{S}$,
with $S=\int dt\, \cal{S}$, and where\footnote{
With our assumptions of static, purely magnetic,  fields we have
an energy density $T_0\sp0={\cal L}_M$, these being the final three terms of
(\ref{redaction}). }
\begin{equation}
\begin{split}
{\cal S}=\int d^{3}x\sqrt{h}
\bigg[ &\frac{1}{16\pi G}\left( R\sp{(3)}(h) +2 \nabla\sp{i}\nabla_{i} \ln V
       -2 \nabla\sp{i}\ln V\: \nabla_{i} \ln V \right)\\
 &-\frac{1}{4}V\sp2 \Tr F_{ij}F^{ij} 
 -\frac{1}{2} \Tr  D_{i}n D^{i} n
+ \frac{\lambda V\sp{-2}}{2}\left(\Tr n\sp2-1\right)
\bigg].
\label{redaction}
\end{split}
\end{equation}
In this reduced action, and indeed from here onwards,
all covariant derivatives are constructed from
and indices lowered {\it etc.} with respect to the three metric $h_{ij}$,
and $R\sp{(3)}(h)$ is the scalar curvature of this metric. We obtain
(\ref{redaction}) by expressing $R(g)$ in terms of $R\sp{(3)}(h)$ and
the function $V$. For the reduced action the equations of motion for $V$
correspond to the $(00)$-Einstein equations (\ref{eins}).

Our strategy to derive the Bogomolny equations is in the spirit of \cite{bogo}
where actions were expressed as sums of squares. First we show that (with an
appropriate choice of units) the reduced action may be expressed in the form
\begin{equation}
{\cal S}=  \int d^{3}x  \sqrt{h}
\bigg[ \frac{1}{4} R\sp{(3)}(h)
-\frac{1}{2} \Tr h\sp{ij} v_i\sp\pm  v_j\sp\pm  \bigg] +{\cal B}\sp\pm ,
\label{newredaction}
\end{equation}
where $v_i\sp\pm$ is an appropriate combination of the fields and
${\cal B}\sp\pm $ is a surface term that will be specified shortly.
The key point is
that the action is quadratic in $v\sp\pm$ (and this quadratic is of definite
signature say for a semisimple gauge group). Thus for any field variation
this term of the action has an extremal when $v_{i}^{a\pm}=0$ for every $i$ and
gauge index $a$. These will be our Bogomolny equations. Further analysis
will then show the three-metric $h_{ij}$ flat.

Towards establishing (\ref{newredaction}) let us implement the Lagrange 
multiplier constraint so that $\Tr n^2=1$. Then
\begin{equation}
\Tr n D_{i}n=0.
\label{nulprod}
\end{equation}
Let $\tau_{ijk}=\sqrt{h}\, \epsilon_{ijk}$ where $\epsilon_{ijk}$ is the
Levi-Civita symbol (satisfying $\epsilon_{123}=+1$). Then the volume
form is $\tau_h =\sqrt{h}\, d\sp3 x=\tau_{ijk}dx^i dx^j dx^k/3!$
and the Hodge star operation is given in terms of $\tau_{ijk}$. For example
$(*F)_i =\tau_{ijk} F\sp{jk}/2$ for a two form $F$. Upon
using the Bianchi identity, the $\sigma$-model constraint and (\ref{nulprod})
one may verify that
\begin{equation}
\begin{split}
\sqrt{h} & \Tr h\sp{ij}
\left[ \frac{1}{2}\tau_{ikl} e\sp{u} F\sp{kl}\pm (n\partial_iu +D_i n)\right]
\left[ \frac{1}{2}\tau_{jmn} e\sp{u} F\sp{mn}\pm (n\partial_ju +D_j n)\right]
\\
&= \sqrt{h}\left( \frac{1}{2}e\sp{2u} \Tr F_{ij}F^{ij}+\Tr  D_{i}n D^{i} n
+\nabla\sp{i}u \nabla_{i} u \right)
\pm \partial_i \Tr\left( \sqrt{h} e\sp{u} \tau\sp{ijk} F_{jk} n\right). \\
\end{split}
\label{factor}
\end{equation}
A comparison of (\ref{factor}) with (\ref{redaction}) shows that we may
rewrite the reduced action in the form (\ref{newredaction}) upon setting
\begin{equation}
V(x\sp1,x\sp2,x\sp3)=e\sp{u},
\label{defV}
\end{equation}
and choosing units so that $ 4\pi G=1$. We have defined $v_i\sp\pm$ in 
(\ref{newredaction}) by
\begin{equation}
v_i\sp\pm = \frac{1}{2}\tau_{ikl} e\sp{u} F\sp{kl}\pm (n\partial_iu +D_i n),
\label{defv}
\end{equation}
while the surface term ${\cal B}\sp\pm $ is
\begin{equation}
{\cal B}\sp\pm =
\frac{1}{2} \int d^{3}x
\partial_i\left( \sqrt{h} h\sp{ij}\partial_j u \pm
\Tr\left( \sqrt{h} e\sp{u} \tau\sp{ijk} F_{jk} n\right) \right). 
\label{defbound}
\end{equation}
Equations (\ref{newredaction}, \ref{defv})
may be succinctly written in terms of differential forms as
\begin{equation}
{\cal S}=  \int \bigg[ \frac{1}{4} R\sp{(3)}(h) \tau_h -\frac{1}{2} \Tr 
v\sp\pm \wedge * v\sp\pm  + \frac{1}{2}
d \left(  *du \pm \Tr(2 e\sp{u} F n) \right) \bigg],
\label{formaction}
\end{equation}
with
\begin{equation}
v\sp\pm=v_i dx\sp{i}= *e\sp{u} F \pm (n du +D n).
\label{formv}
\end{equation}
We remark that the requirement $ 4\pi G=1$ sets a mass scale for the model.
It has the same effect as in classical Newtonian mechanics of 
allowing static configurations of self-gravitating point charges when the
masses and charges agree in suitable units. It is analogous to the choice
of a  critical coupling when constructing Bogomolny equations for the
Abelian-Higgs model of superconductors.

As we commented upon earlier, the quadratic  appearance of $v\sp\pm$
in the action  means that this term has an extremal for any field
variation when $v_{i}^{a\pm}=0$ (for every $i$ and
gauge index $a$). These are the desired Bogomolny equations.
In particular, supposing $v_{i}^{a\pm}=0$,  then varying the
action with respect to the three-metric $h_{ij}$ yields,
\begin{equation*}
\delta \int d^{3}x \sqrt{h}\; R^{(3)}(h)\;=0.
\end{equation*}
This shows $h_{ij}$ is a solution of the three dimensional  Einstein equations
\begin{equation*}
R^{(3)}_{ij}(h)-\frac{1}{2}h_{ij}R^{(3)}(h)=0,
\end{equation*}
which in turn implies 
\begin{equation*}
\begin{array}{cc}
R^{(3)}(h)=0,&R^{(3)}_{ij}(h)=0.
\end{array}
\end{equation*} 
Using the three dimensional identity
\begin{equation*}
R_{ijkl}=h_{ik} R_{jl} - h_{il} R_{jk} + h_{jl} R_{ik} - h_{jk} R_{il} 
         +\frac{1}{2}( h_{il}  h_{jk} -h_{ik}  h_{jl})R
\end{equation*} 
we deduce that the three-metric $h_{ij}$ is necessarily flat.
This shows the various flat models assumed by previous authors were
in fact necessary.
If space is topologically ${\mathbb R}\sp3$ then this flatness means it 
is isometric to Euclidean space and we may take $h_{ij}=\delta_{ij}$, 
in which case the four-metric is
\begin{equation}
ds\sp2 = -V\sp2 dt\sp2 +\frac{1}{V\sp2} d{\bf x}\cdot d{\bf x}.
\label{isometric}
\end{equation}

Let us summarise our calculations thus far. Assuming $n$ satisfies
$\Tr n^2=1$  (and with a choice of units such that $ 4\pi G=1$) we find 
\begin{equation}
n\sp{a}\partial_i u +D_i n\sp{a} =\mp 
\frac{1}{2}\tau_{ikl} e\sp{u} F\sp{a\,kl},
\label{Bogomolny1}
\end{equation}
or equivalently
\begin{equation}
n du +D n = \mp *e\sp{u} F,
\label{Bogomolny2}
\end{equation}
yield solutions of the field equations (\ref{ym}-\ref{eins}) if and only
if the three-metric $h_{ij}$ defined by (\ref{metric}) is flat. 
These are our Bogomolny equations for the action (\ref{action}); one can
readily begin with (\ref{Bogomolny1}) or (\ref{Bogomolny2}) and a flat
three-metric and derive (\ref{ym}-\ref{eins}).
We have then established our first result and shown the compatibility
of the Bogomolny equations with the remaining Einstein equations, so
justifying the ansatz of the various authors mentioned in the introduction.

As we mentioned in the introduction, the spacetimes with metric
(\ref{isometric}) are precisely those  static space-times for which 
super-covariantly constant spinors  can exist \cite{gibbonshull}, and
Tod \cite{tod} has extended this classification to stationary metrics.
Hlousek and Spector \cite{HlSpect} have advanced arguments associating
supersymmetric extensions to theories exhibiting Bogomolny bounds.
This suggests a supersymmetric extension exists to the theory here
under consideration. Here we will neither pursue the construction of such a 
theory nor address the issue of whether super-covariantly constant spinors
exist; the recent work of \cite{gum} looks at various 4 dimensional
supergravity reductions to sigma models addressing related issues.

\section{Solutions}
It remains to discuss solutions of the  Bogomolny equations (\ref{Bogomolny1}, %
\ref{Bogomolny2}). Before focusing on the case of 
axially symmetric solutions for the gauge group $SU(2)$ we first show that 
an auxiliary  magnetostatic problem \footnote{Equally one may
consider this to be an auxiliary electrostatic problem. Because we have in
mind the nonabelian problem where the sources are magnetic monopoles,
we adopt the magnetostatic perspective.}
may be associated with the equations and  then discuss the 
case of covariantly constant solutions.

\subsection{General Properties}
First then let us consider projecting the Bogomolny equations in the
direction of the $\sigma$-model field. This is analogous to projecting the 
usual Bogomolny equations of Yang-Mills-Higgs theory in the 
direction of the Higgs fields. Using  (\ref{nulprod}) we find
\begin{equation}
\partial_i (e\sp{-u})  =\pm
\frac{1}{2}\tau_{ikl} \Tr( F\sp{kl} n),
\label{projBogmolny1}
\end{equation}
(equivalently $de\sp{-u}=\pm *\Tr(F n)$) and  this together with the Bianchi
identities yields
\begin{equation}
\nabla\sp{k}\nabla_k e\sp{-u}=\pm \frac{1}{2}\tau\sp{ijk} \Tr( F_{ij}
D_k n).
\label{poisson}
\end{equation}
Upon setting $b_i=\partial_i (e\sp{-u})$ and 
$\vec{b}=\left(b_{1},b_{2},b_{3}\right)$ we may recast our equations
in the form
\begin{equation}
\nabla \cdot \vec{b}=\rho,\quad
\nabla\times\vec{b}=0,\quad
\nabla^{2}\Phi=\rho.
\label{magnetostatic}
\end{equation}
Here
\begin{equation}
\rho=\pm \frac{1}{2}\tau\sp{ijk} \Tr( F_{ij} D_k n),\quad
\vec{b}= \nabla\Phi,\quad
\Phi=e\sp{-u}.
\label{magnetodefs}
\end{equation}
We see then from (\ref{magnetostatic}) that there 
is an Abelian magnetostatic problem associated with (\ref{Bogomolny1})
in which $\Phi$, $\vec{b}$, and $\rho$ play the role of the magnetic scalar 
potential, the magnetic field and the magnetic charge density, respectively.
In particular the original metric coefficient $g_{00}$ is entirely determined
by the Poisson equation of (\ref{magnetostatic}).

We observe that for the special sub-class of
covariantly constant scalar solutions, those characterised by
\begin{equation} D_{i}n^{a}=0,
\label{covconst}
\end{equation}
the projected Bogomolny equations (\ref{projBogmolny1}) in fact encode
the full Bogomolny equations. In this situation the magnetic charge density
$\rho$ (the source in the Poisson equation above) vanishes and the
metric coefficient $g_{00}$ is determined by the solution of a Laplace
equation.
Further, $\vec{b}$ is
divergenceless in this particular situation and 
so it is the curl of an Abelian vector potential which has
string singularities defined in the associated
three-dimensional Euclidean space. The general solution for $V$ in this 
situation describes a system of
point-like monopoles in static equilibrium interacting with the
gravitational field. These are the Majumdar-Papapetrou solutions 
\cite{maj, pap}.  With only one monopole placed at the
origin of coordinates this (positive mass) solution takes the form 
\begin{equation}
e\sp{u}=\frac{1}{1+\frac{|Q|}{r}},
\label{solV}
\end{equation}
where $Q$ is the magnetic charge and 
$r=\sqrt{(x^{1})^{2}+(x^{2})^{2}+(x^{3})^{2 }}$. 
The asymptotic form of the metric is given by
\begin{equation}
|g_{00}|\approx 1-\frac{2 |Q|}{r},
\label{asym1}
\end{equation}
with a total gravitational mass $M=4\pi |Q|$.
Such a solution corresponds to  an extreme Reissner-Nordstr\"{o}m black hole
in which the null hypersurface $r=0$ corresponds to the horizon with radius
$|Q|$ in Schwarzschild coordinates.

\subsection{$su(2)$ Solutions}
In order to proceed further with our analysis of the Bogomolny
equations (\ref{Bogomolny1}) we restrict our attention to the $su(2)$ case.
When the scalar field is non-covariantly constant the full nonlinearity
of the Bogomolny equations becomes apparent. In
order to solve these equations we start by constructing a static,
axially symmetric, magnetic ansatz for the components of the Yang-Mills 
connection. Upon combining the resulting first order equations we shall obtain 
a single second order partial differential equation (\ref{pdechi}) governing 
the system and from whose solutions the various fields may be reconstructed.
The restriction to spherical symmetry reproduces the ordinary differential
equation obtained by earlier authors.
Unfortunately no analytic solutions for (\ref{pdechi})  (or indeed its
spherically symmetry reduction) are known and we must consider
various approximations.

In the $su(2)$ setting a non-covariantly constant $D_{i}n^{a}$ 
may be expressed as
\begin{equation}
D_{i}n^{a}=\alpha_{i}p^{a}+\beta_{i}q^{a},
\label{notrivder}
\end{equation}
where $\alpha_{i}$ and $\beta_{i}$ are functions of the space
coordinates and $p^{a}$ and $q^{a}$ are chosen to satisfy
$p^{a}n^{a}=q^{a}n^{a}=0$, so that (\ref{nulprod}) holds. Imposing the
additional conditions $p^{a}p^{a}=q^{a}q^{a}=1$ and
$q^{a}=\epsilon_{abc}n^{b}p^{c}$, the triad $(n,p,q)$ becomes a
rotating, orthonormal base for $su(2)$. 
Working in spherical coordinates, we adopt an ansatz in which
we assume that $n$, $p$, and $q$ take the special forms
\begin{equation}
\begin{array}{l}
n=\sin\theta\cos\phi\; T_{1}+\sin\theta\sin\phi\; T_{2}+\cos\theta\; T_{3}
\equiv \hat x\sp{a} T_a ,\\
p=\cos\theta\cos\phi\; T_{1}+\cos\theta\sin\phi\; T_{2}-\sin\theta\; T_{3}
,\\
q=-\sin\phi\; T_{1}+\cos\phi\; T_{2},
\end{array}
\label{npq}
\end{equation}
where $T_{1},T_{2},T_{3}$ are the generators of the adjoint
representation of $su(2)$.\footnote{ With $[T_a,T_b]=\epsilon_{abc}T_c$ we have
$$q=[n,p],\quad n=[p,q],\quad p=[q,n],$$
and
$$ dn= p\, d\theta +\sin\theta  q\, d\phi,\quad
   dp= -n\,  d\theta +\cos\theta q\, d\phi,\quad
   dq=-(\sin\theta n+\cos\theta p)d\phi.
$$
}

Now upon combining the definition of the covariant derivative $D_{i}n^{a}$ 
with (\ref{notrivder}) we may solve (extending \cite{cofn}) 
for the Yang-Mills potential
$A=A_{i}\;dx^{i}=A^{a}_{i}\;T_{a}\;dx^{i}$, to obtain 
\begin{equation}
A^{a}_{i}= \alpha_{i}q^{a}-\beta_{i}p^{a}-\epsilon_{\ bc}\sp{a}
n^{b}\partial_{i}n^{c} -\delta_{i}n^{a},
\label{ans1}
\end{equation}
where $\delta_{i}$ are three arbitrary functions of $x^{i}$. 
This means the components of the gauge field may be expressed in polar
coordinates as
\begin{equation}
A^{a}=q^{a} 
(\alpha-d\theta)+p^{a}(\sin\theta d\phi
-\beta)- n^{a}\delta .
\label{Acomps}
\end{equation}
By projecting (\ref{Bogomolny2}) in the $(n,p,q)$ directions we find the
equations connecting these various unknowns ($\alpha= \alpha_{i}\;dx^{i}$,  
{\it etc.}) to be
\begin{equation}
\begin{array}{rl}
du&= \mp\star e\sp{u}\left(
-d\delta +\alpha\wedge \beta -\sin\theta d\theta\wedge d\phi 
\right),\\
\alpha&= \mp\star e\sp{u}\left(
-d\beta+\delta\wedge\alpha+\alpha\wedge\cos\theta d\phi 
\right),\\
\beta&= \mp\star e\sp{u}\left(
d\alpha-\beta\wedge\delta+\beta\wedge\cos\theta d\phi
\right).
\end{array}
\label{Bogform}
\end{equation}
Further, we have that
\begin{equation}
\begin{array}{cc}
\left[L_{3}- T_{3},A\right]= q\, \partial_\phi \alpha -p\, \partial_\phi \beta
-n\, \partial_\phi \delta,
\end{array}
\label{axsygen}
\end{equation}
where $L_{3}$ is the generator of space-rotations around the
$x^{3}$ axis.  By taking $\alpha$, $\beta$ and $\delta$ independent of
$\phi$ we then have
\begin{equation}
\begin{array}{cc}
\left[L_{3}- T_{3},A_{i}\right]=0.
\end{array}
\label{axsycon}
\end{equation}
This property \footnote{
It is perhaps worth remarking that an axially symmetric connection
does not necessitate an axially symmetric space-time. This will occur
when $V$ (or equivalently $e\sp{u}$) is independent of $\phi$. From
(\ref{Bogform})  we see this means
$$
0=\partial_r \delta_{2} -\partial_\theta \delta_{1} -\alpha_{1}\beta_{2}
+\alpha_{2}\beta_{1},
$$
and so for (\ref{metric}) to admit 
${\partial}/{\partial\phi}$ as a Killing vector implies
a constraint on some of the connection parameters.
Similarly further constraints are needed for the metric to be
spherically symmetric.}
is characteristic of an axially symmetric, $su(2)$ connection \cite{rebi}.

At this stage we have expressed the Bogomolny equations (\ref{Bogomolny1}) 
for the gauge group $su(2)$ in the form (\ref{Bogform}). No approximations
have so far been introduced and the problem remains of either solving
(\ref{Bogform}) or finding ansatz that enable their solution.

Our ansatz for the study of (\ref{Bogomolny1}) is now
based on the following choice for
$\alpha_{i}$, $\beta_{i}$, and $\delta_{i}$:
\begin{equation}
\begin{array}{lll}
\alpha_{1}=0,&\alpha_{2}=\chi,&\alpha_{3}=0,\\
\beta_{1}=0,&\beta_{2}=0,&\beta_{3}=\sin\theta\;\chi,\\
\delta_{1}=0,&\delta_{2}=0,&\delta_{3}=\psi,
\end{array}
\label{abc}
\end{equation}
where $\chi$ and $\psi$ are in general functions of $r$ and $\theta$. 
These choices mean (\ref{axsycon}) is satisfied and we have an 
axially symmetric, $su(2)$ connection.
Spherically symmetric solutions are covered within our
ansatz by $\psi=0$, $\chi=\chi(r)$, in which case we recover the t'Hooft ansatz,
$$
A\sp{a}_i = \epsilon\sp{a}_{\  ib}\, \frac{\hat x \sp{b} }{r} (1-\chi(r)).
$$

With this ansatz and taking\footnote{
Choosing the opposite sign leads to the same class of solutions.
The sign ambiguity can be absorbed by working with a quantity $\pm e\sp{u}$;
this sign  is responsible for the existence of solutions with either positive
or negative magnetic charge {\it and} positive gravitational mass.}
the (-) sign in (\ref{Bogomolny1})
we find five distinct equations amongst the components of the Bogomolny 
equations (\ref{Bogform}):
\begin{equation}
\frac{\partial}{\partial r}\chi=- e\sp{-u} \chi
\label{bogop1}
\end{equation}
\begin{equation}
\frac{\partial}{\partial \theta}\chi=-\frac{\psi\chi}{\sin\theta},
\label{bogop2}
\end{equation}
\begin{equation}
\frac{\partial}{\partial r}\left(e\sp{-u}
\right)=-\frac{1}{r^{2}}\left(1-\chi^{2}+\frac{1}{\sin\theta}\frac{\partial 
\psi}{\partial \theta}\right),
\label{bogop3}
\end{equation}
\begin{equation}
\frac{\partial}{\partial \theta}\left(e\sp{-u}
\right)=\frac{1}{\sin\theta}\frac{\partial \psi}{\partial r},
\label{bogop4}
\end{equation}
\begin{equation}
\frac{\partial}{\partial \phi}\left(e\sp{-u}
\right)=0.
\label{bogop5}
\end{equation}
Not all of these equations are independent:
(\ref{bogop4}) is a consequence of (\ref{bogop1}) and
(\ref{bogop2}).
Now equation (\ref{bogop5}) implies that our conformstatic metric
(\ref{metric}) admits $\frac{\partial}{\partial\phi}$ as a Killing
vector.
Within our ansatz the $\sigma$-model equations (\ref{sf})
reduce to (\ref{bogop2}) and  the axisymmetric charge density
$\rho$ (\ref{magnetodefs}) simplifies to
\begin{equation}
\rho=\frac{1}{r^2}\frac{\partial(\chi^{2})}{\partial r}.
\label{exprerho}
\end{equation}
The latter result indicates that $e\sp{-u}=\frac{1}{V}$ 
 is  not in general  an harmonic function.
Combining (\ref{bogop1}-\ref{bogop3})
yields the equation \begin{equation} r^{2}\frac{\partial^{2}}{\partial r^{2}}\ln
|\chi|+\frac{1}{\sin\theta}\frac{\partial}{\partial\theta}\left(\sin\theta\:\frac{\partial}{\partial\theta}\ln|\chi|\right)=1-\chi^{2}.
\label{pdechi}
\end{equation}
This equation governs the solutions of our Bogomolny equations.
Once we have a solution for (\ref{pdechi}), the remaining unknowns $V$ and 
$\psi$ can be simply determined using (\ref{defV}),
(\ref{bogop1}) and (\ref{bogop2}).  

At this stage we have obtained the single second order partial differential 
equation (\ref{pdechi}) governing the system.
In the case of spherical symmetry this equation reduces to a form
of Emden's isothermal gas equation. (In this case the $\sigma$-model equations 
(\ref{sf}) become trivial as (\ref{bogop2}) vanishes.) With
$\ln|\chi(r)|=-F(r)-\ln r$ we obtain equation (21) of \cite{cho},
\begin{equation}
r\sp4 F'' =e\sp{-2 F}.
\label{emden}
\end{equation}
Such Emden type equations may be rewritten in the form of an Abel
equation
\begin{equation}
\frac{d \chi}{d\bar r}= -\frac{\bar r \chi}{\bar r -1 +\chi\sp2},
\label{abel}
\end{equation}
where $\bar r= r\, F'(r)-1=r\, e\sp{-u}$ (and $\chi=e\sp{-F}/r$).
Indeed the variable $\bar r$ naturally arising in this transformation
is such that the metric takes the form
\begin{equation}
ds\sp2= -e\sp{2u} \, dt\sp2 +e\sp{-2u}\left(\frac{dr}{d\bar r}\right)\sp2
{d\bar r} \sp2+ {\bar r}\sp2 (d\theta\sp2 +\sin\sp2 \theta d\phi\sp2).
\label{schwmetric}
\end{equation}
Unfortunately no analytic solutions for (\ref{pdechi})  (or indeed the
spherically symmetry reductions (\ref{emden}) or (\ref{abel})) are known and 
we must consider various approximations.
The form of the metric (\ref{schwmetric}) is helpful when considering 
possible singularities of solutions, with $4\pi{ \bar r}\sp2$ being the
area of a two-sphere about the centre of symmetry. We see from this that the
conditions for a horizon are 
\begin{equation}
r \frac{du}{d r}\bigg|_{H} =1,   \qquad e\sp{u}\bigg|_{H} =0,
\label{bhbcs}
\end{equation}
and consequently $r=0$ at a horizon. 

It follows from our equations that an asymptotically flat, spherically 
symmetric black hole solution with regular gauge fields is necessarily
an extreme Reissner-Nordstr\"{o}m black hole. This may be argued as 
follows.\footnote{This argument is based on one shown to us by
E.J. Weinberg and D. Maison.}
The spherically symmetric Bogomolny 
equations together with (\ref{bhbcs}) entail $\chi\big|_{H} =0$ for regular 
gauge fields.
Further, the asymptotic boundary condition on the gauge field is 
$\chi\big|_{\infty}=0$. Now it follows from (\ref{bogop1}) that
$\chi$ is a monotone function from the horizon to radial infinity,
and so $\chi(r)$ vanishes identically. This brings us to the covariantly
constant extreme Reissner-Nordstr\"{o}m solution of the previous section.
This result agrees with the work of \cite{BFM} who have analysed the
region of parameters for which spherically symmetric EYMH black holes exist:
our sigma-model corresponds to their ($\beta=\infty$) infinite mass limit,
while the Bogomolny equations necessitate their remaining parameter to be
$\alpha=1$.
Without the restriction of spherical symmetry it is more difficult to determine
the singularity structure of our solutions. The 
Majumdar-Papapetrou solutions of the previous section may be shown to have 
singularities shielded by horizons. We will defer such discussion on the 
nature of singularities until the appendix, looking here to the
possible asymptotics of solutions.

If $|\chi (r)|$ is small for large $r$, then (\ref{pdechi})
and (\ref{bogop1}) have the approximate (asymptotically flat) solution 
\begin{equation}
\begin{array}{cc}
\chi(r)\approx B\:\frac{e^{-r}}{r},&
e\sp{u}\approx 1-\frac{1}{r},
\end{array}
\label{appse}
\end{equation} 
where $B$ is an arbitrary constant. Comparing this asymptotic solution with
(\ref{asym1}) we conclude that (\ref{appse}) corresponds to a magnetic
monopole with unit magnetic charge and total gravitational mass
$M=4\pi$.
Another approximate solution for (\ref{pdechi}) is
\begin{equation}
\ln |\chi|\approx C\sqrt{r}\sin\left(\frac{\sqrt{7}}{2}\ln r +\Omega\right),
\label{solbawa}
\end{equation} 
where $C$ and $\Omega$ are integration constants. As discussed by
Balakrishna and Wali \cite{bawa}, (\ref{solbawa}) is valid for
$|\chi|\approx 1$, which is the case when $r\rightarrow 0$. The
oscillatory behaviour of (\ref{solbawa}) implies a countable, infinite
set of (naked) singular spheres surrounding the origin, which is the
locus of an event horizon of zero surface area. (See \cite{bawa} for
details.) 
We have found numerical evidence for the existence of asymptotically flat 
solutions of (\ref{pdechi}) which smoothly interpolate  between the 
(asymptotic) monotone and oscillatory regimes described approximately by 
(\ref{appse}) and (\ref{solbawa}), a point left unanswered in earlier works.
This is described in the appendix
alongside remarks on the nature of the singularities associated to 
such solutions.

In the non-spherically symmetric case, we may begin with the
approximation (for large $r$) to (\ref{pdechi}),
\begin{equation} 
\chi\approx B\:
\frac{e^{-r}}{r}\left[1+\epsilon
\sum\limits_{l=1}^{\infty}\frac{f_{l}}{r^{l}}\:P_{l}(\cos\theta)\right].
\label{asypaxi}
\end{equation}
Here $B$ is is arbitrary and $\epsilon$ is a small parameter.
Upon using (\ref{bogop1}) and dropping terms of order $\epsilon\sp2$ we find
\begin{equation}
e\sp{u}\approx 1-
\left[\frac{1}{r}+ \epsilon
\sum\limits_{l=1}^{\infty}\frac{l\:f_{l}}{r^{l+1}}\:P_{l}(\cos
\theta)\right].
\label{axiVasymp}
\end{equation}
Comparing with (\ref{defV}) we see this corresponds to an asymptotically 
flat solution.
Similarly using (\ref{bogop2}) we  obtain
\begin{equation}
\psi\approx -\epsilon
\sin\theta\sum\limits_{l=1}^{\infty}\frac{f_{l}}{r^{l}}\:
\frac{dP_{l}(\cos\theta)}{d\theta}.
\label{axipsi}
\end{equation}
Observe that, in contrast to $\chi$,
$\psi$ is a long range potential. It is also possible to find an
approximate solution of (\ref{pdechi}) for small $r$, such that $\chi$
is finite at $r=0$: 
\begin{equation} |\chi|\approx 1+C\sqrt{r}\sin\left(
\frac{\sqrt{7}}{2}\ln r+\Omega\right)+
\sum\limits_{l=1}^{\infty}c_{l}\:r^{\frac{1}{2}+\sqrt{l(l+1)-\frac{7}{4}}}
\:P_{l}(\cos\theta),
\label{axichi}
\end{equation}
where $C$, $\Omega$, and $c_{1}, c_{2}, c_{3},....$ are arbitrary constants. 
This expression shows a remarkable difference between the radial dependence of 
the monopole and higher multipole terms in $\chi$. We see that 
(\ref{asypaxi}-\ref{axichi}) reduce to the previous spherically-symmetric 
results when every $f_{l}$ and $c_{l}$ vanishes.

A simple perturbative method can be used in order to improve the
asymptotic solutions (\ref{appse}) and
(\ref{asypaxi}-\ref{axipsi}). We illustrate the procedure in the
spherically symmetric case only, but it can be extended
straightforwardly to the case in which $\chi$ depends on $\theta$ as
well. 
We have for large $r$ that $\chi (r)\approx B{e^{-r}}/{r}$ which gives the
correct asymptotic behaviour for the metric. Now suppose that
$$
\chi (r)=B\frac{e^{-r}}{r}\left(1+\varepsilon\right)
$$
where $\varepsilon\ll 1$. Upon substituting in (\ref{pdechi})
we find the approximate expression
\begin{equation}
\varepsilon=-\frac{B^{2}}{4}\frac{e^{-2r}}{r^{4}}.
\label{varep}
\end{equation} 
As a consequence, our new asymptotic solutions are
\begin{equation}
\chi\approx
B\frac{e^{-r}}{r}\left(1-\frac{B^{2}}{4}\:\frac{e^{-2r}}{r^{4}}\right),
\end{equation} 
\begin{equation} 
|g_{00}|\approx 1-\frac{2m(r)}{r}, 
\label{improvedg00}
\end{equation} 
where 
\begin{equation}
m(r)=1-\frac{B^{2}}{2}\frac{e^{-2r}}{r^{3}}.
\label{mrimproved}
\end{equation} 
This improved asymptotic approximation exhibits a very small
correction in the distribution of gravitational mass, suggesting the
existence of a massive, extended magnetic core.
Such exponential corrections also  arise from instanton corrections.

\section{Inclusion of a Dilaton}
We now extend our earlier analysis to include a dilaton. Up to
a redefinition of $u$ we will find that the Bogomolny equations are
unaltered. Consider the EYMD$\sigma$ system  given by the action
\begin{equation}
\begin{split}
S_d=\int d^{4}x\sqrt{|g|}
\bigg[&\frac{1}{16\pi G}\left(R-\frac{1}{2}\partial_{\mu}\varphi\partial^{\mu}
\varphi\right)
 - \frac{1}{4} \Tr \left(e^{\gamma\varphi}F_{\mu\nu}F^{\mu\nu}\right)\\
&-\frac{1}{2} \Tr \left( D_{\mu}nD^{\mu}n\right)
+\frac{\lambda}{2}\left(\Tr n\sp2-1\right)
\bigg].
\end{split}
\label{daction}
\end{equation}
The field equations are now (\ref{sf}) and (\ref{eins}) together with the
modified gauge equation
\begin{equation}
\frac{1}{\sqrt{|g|}}D_{\mu}\left(\sqrt{|g|}e^{\gamma\varphi}F^{\mu\nu}\right)
=\left[n,D^{\nu}n\right],
\label{ymd}
\end{equation}
and the new dilaton field equation
\begin{equation}
\frac{1}{\sqrt{|g|}}\partial_{\mu}\left(\sqrt{|g|}\partial^{\mu}\varphi\right)
= \left(4\pi G \gamma \right)e^{\gamma\varphi}\Tr F_{\mu\nu}F^{\mu\nu}.
\label{dil}
\end{equation}
Upon making use of (\ref{ric00}) one observes \cite{forhorhor} that
with the identification
\begin{equation}
\varphi=2\gamma\; \ln\sqrt{|g_{00}|}
\label{sigmacon}
\end{equation}
the dilaton equation of motion (\ref{dil}) coincides with the $(00)$-Einstein
equation. Our construction of a
reduced action now proceeds as before. To employ the
identity (\ref{factor}) we want
$$
e\sp{2u}=V\sp{2} e^{\gamma\varphi},
$$
which, upon using (\ref{sigmacon}), requires
\begin{equation}
u=\left(1+\gamma^{2}\right)\ln\sqrt{|g_{00}|}\;; 
\quad i.e.\quad
e\sp{u}=V\sp{1+\gamma^{2}}.
\label{def1}
\end{equation}
Finally, the choice
\begin{equation}
4\pi G\left(1+\gamma^2\right)=1
\label{unitsd}
\end{equation}
means we have a reduced action 
\begin{equation}
{\cal S}_d=  \int d^{3}x  \sqrt{h}
\bigg[ \frac{1+\gamma^{2}}{4} R\sp{(3)}(h)
-\frac{1}{2} \Tr h\sp{ij} v_i\sp\pm  v_j\sp\pm  \bigg] +{\cal B}_d\sp\pm ,
\label{newredactiond}
\end{equation}
where ${\cal B}_d\sp\pm $ is again a surface term and $v_i\sp\pm$ is given
by (\ref{defv}). Up to the scaling of the scalar curvature this is of
an identical form to (\ref{newredaction}). In particular this means we obtain 
the same Bogomolny equations as previously with the same conclusion that the
three-metric is flat.

Therefore, assuming $n$ satisfies $\Tr n\sp2=1$ and with a choice of
units (\ref{unitsd}), we find that the Bogomolny equations 
(\ref{Bogomolny1}-\ref{Bogomolny2}) 
again provide solutions to the field equations for
(\ref{daction}) if and only if the three-metric $h_{ij}$ is flat.
The dilaton field is given by (\ref{dil}) for such solutions.

An important consequence of having the same Bogomolny equations is that the
solutions discussed in the previous section may be directly used in the
present setting, though with the metric function $V$ now related to
$e\sp{u}$ via  (\ref{def1}) rather than (\ref{defV}).
Thus in the absence of a dilaton the solution (\ref{solV}) to Laplace's
equation describing a single monopole led to the asymptotic form of the
metric (\ref{asym1}) we now find
\begin{equation*}
\begin{array}{cc}
|g_{00}|\approx 1-\frac{2m}{r},& m=\frac{|Q|}{1+\gamma^{2}}\:.
\end{array}
\end{equation*}
Whereas without the dilaton we have an extreme  Reissner-Nordstr\"{o}m black 
hole, according to Ref. \cite{cho} the solution with dilaton has a naked, 
point-like singularity at $r=0$ for $\gamma \neq 0$.
The energy integral may be determined 
analytically for this solution with arbitrary $\gamma$ yielding
\begin{equation*}
E=\int d^{3}x \sqrt{|g|}\; T^{0}{}_{0}=
\frac{2\pi (1+2\gamma^{2})|Q|}{1+\gamma^{2}}\;.
\end{equation*}
The fact that E is finite in the case $\gamma\neq 0$ -in which the integration 
region includes the naked singularity-  can be considered as the
extension to curved spacetime of a result previously found by
Bizon \cite{bizon}, who obtained a
Bogomolny-type solution for the Yang-Mills-dilaton equations.
(See also \cite{lavrelashvili} for an independent discussion of these 
equations.)
Park \cite{park} has also discussed the naked singularity structure
of dilatonic point-like Taub-NUT multi-body solutions to an 
Einstein-Maxwell-Dilaton theory.

Similarly with our axially symmetric solutions the
metric function $V$ is modified accordingly, based on the same function 
$e\sp{u}$.
Thus to a first approximation (\ref{appse}) leads to
$$
V\approx 1-\frac{1}{1+\gamma^{2}}\frac{1}{r}
$$
which may be improved via (\ref{improvedg00}) with
$$
m(r)=\frac{1}{1+\gamma^{2}}\left[1-\frac{B^{2}}{2}\frac{e^{-2r}}{r^{3}}\right].
$$

\section{Solutions with Higher Winding Numbers}

In this Section we study axisymmetric solutions of the modified Bogomolny 
equations with winding number greater than one.
Again the problem lies in finding tractable ansatz for the exact
Bogomolny equations.
To this end, let us consider the following \cite{kleihaus, rebi}  magnetic,
static prescription for the SU(2) potentials:
\begin{equation}
A_{\mu}\;dx^{\mu}=q^{(k)}\left[-\frac{H_{1}}{r}dr+\left(H_{2}-1\right)d\theta
\right]+k\left[n^{(k)}\;H_{3}+p^{(k)}\left(1-H_{4}\right)\right]\sin\theta 
d\phi,
\label{ans2}
\end{equation}
where $n^{k}$, $p^{k}$ and $q^{k}$ are defined by
\begin{equation}
\begin{array}{l}
n^{(k)}=\sin\theta\cos k\phi\; T_{1}+\sin\theta\sin k\phi\; T_{2}+\cos\theta\; 
T _{3},\\
p^{(k)}=\cos\theta\cos k\phi\; T_{1}+\cos\theta\sin k\phi\; T_{2}-\sin\theta\; 
T _{3}
,\\
q^{(k)}=-\sin k\phi\; T_{1}+\cos k\phi\; T_{2},
\end{array}
\label{npqk}
\end{equation}
and the four functions $H_{i}$ depend on $r$ and $\theta$ only.
The integer $k$ will be interpreted as the winding number of the solutions. 
(Equations (\ref{ans2}) and (\ref{npqk}) have been inspired by the ansatz 
considered 
in \cite{kleihaus}, which is set up in the fundamental representation of SU(2).)
This new prescription for the connection can also be obtained as a consequence 
of (\ref{ans1}) if we replace\footnote{
Now
$ dn^{(k)}= p^{(k)}\, d\theta +k\,\sin\theta  q^{(k)}\, d\phi$,
$ dp^{(k)}= -n^{(k)}\,  d\theta +k\,\cos\theta q^{(k)}\, d\phi$,
$ dq^{(k)}=-k\,(\sin\theta n^{(k)}+\cos\theta p^{(k)})d\phi$ and
$A=q^{(k)} (\alpha-d\theta)+p^{(k)}(k \sin\theta d\phi -\beta)- n^{(k)}\delta $.
}
$n,\:p,\:q$ by $n^{(k)},\:p^{(k)},\:q^{(k)}$, 
respectively, and make the following choice of parameters:
\begin{equation}
\begin{array}{lll}
\alpha_{1}=-\frac{H_{1}}{r},&\alpha_{2}=H_{2},&\alpha_{3}=0,\\
\beta_{1}=0,&\beta_{2}=0,&\beta_{3}=k\sin\theta\;H_{4}\\
\delta_{1}=0,&\delta_{2}=0,&\delta_{3}=-k\sin\theta\;H_{3}.
\end{array}
\label{abc2}
\end{equation}
This ansatz satisfies the axial symmetry condition
\begin{equation*}
\left[L_{3}-k\,T_{3},A_{i}\right]=0,
\label{axsycon1}
\end{equation*}
which generalises (\ref{axsycon}) to higher winding numbers.
We also observe that, for $k=1$ and $H_{1}=0$, $H_{2}=H_{4}=\chi(r,\theta)$, $H_
{3}=-\frac{\psi(r,\theta)}{\sin\theta}$, our original axially symmetric ansatz 
(\ref{abc}) is recovered.

Combining this new ansatz for the connection with the projections of the 
Bogomolny equations on the rotating basis (\ref{npqk}), we obtain the following
system of equations:
\begin{equation}
{H_{1}}\,{e\sp{-u}}=
\frac{\mp k}{r}\left[\frac{\partial H_{4}}{\partial\theta}+
\cot\theta\:\left(H_{4}-H_{2}\right)-H_{2}H_{3}\right],
\label{b11}
\end{equation} 
\begin{equation}
{H_{2}}\, {e\sp{-u}}=
\frac{\mp k}{r}\left[r\:\frac{\partial H_{4}}{\partial r}
+\cot\theta\:H_{1}+H_{1}H_{3}\right],
\label{b12}
\end{equation}
\begin{equation}
{\mp k\:H_{4}}\,{e\sp{-u}}=\frac{\partial H_{2}}{\partial r}+\frac{1}{r}
\:\frac{\partial H_{1}}{\partial\theta},
\label{b23}
\end{equation}
\begin{equation}
\frac{\partial}{\partial r}\left(e\sp{-u}\right)=
\frac{\mp k}{r^{2}}\left[1-\frac{\partial H_{3}}{\partial\theta}
-\cot\theta\:H_{3}-H_{2}H_{4}\right],
\label{31}
\end{equation}
\begin{equation}
\frac{\partial}{\partial\theta}\left(e\sp{-u}\right)=
\mp k\left[\frac{\partial H_{3}}{\partial r}-\frac{1}{r}\:H_{1}H_{4}\right],
\label{b32}
\end{equation}
\begin{equation}
\frac{\partial}{\partial\phi}\left(e\sp{-u}\right)=0.
\label{b33}
\end{equation}

An asymptotically flat, covariantly constant solution of 
(\ref{b11}-\ref{b33}) 
can be easily obtained if we assume $H_{1}=H_{2}=H_{4}=0$ and
$H_{3}=-\frac{\psi(r,\theta)}{\sin\theta}$. 
The harmonic function $e\sp{-u}$ is given by 
\begin{equation*}
e\sp{-u}=1+\frac{|Q|}{r}+\sum\limits_{l=1}^{\infty}\frac{B_{l}
}{r^{l+1}}\:P_{l}(\cos\theta),
\end{equation*}
where $Q=k$ is the total magnetic charge.
The corresponding solution for $\psi$ is
\begin{equation}
\psi=\psi_{\infty}-\frac{\sin\theta}{k}
\:\sum\limits_{l=1}^{\infty}\frac{C_{l}}{lr^{l}}\:
\frac{dP_{l}(\cos\theta)}{d\theta}.
\label{psik}
\end{equation}
The identification of total
magnetic charge with the winding number in this solution implies that 
the gravitational mass is
conserved for topological reasons. (See \cite{viwa} for a discussion
of this point in the case with no dilaton present.) 
We have yet to solve this ansatz for non-covariantly constant
$\sigma$-model solutions.

\section{Discussion}

This paper has examined the  Bogomolny equations and their solutions
for Einstein-Yang-Mills-$\sigma$ models (with possible dilaton couplings)
on static space-times.
Our derivation of the Bogomolny equations  leads to several new
observations. In particular the Bogomolny equations are consistent with
the Einstein equations if and only if the associated (conformally scaled)
three-metric is flat. These are precisely the  static metrics for which
super-covariantly constant spinors exists, and this class of metrics
includes all of the particular ansatz considered by previous authors.
(Stationary metrics for which super-covariantly constant spinors exist
have also been classified, and the extension of such models to these
space-times will be dealt with elsewhere.)
The connection with super-covariantly constant spinors  suggests a
supersymmetric extension of these models that we have not so far pursued.
Although we have assumed a $\sigma$-model interaction throughout we do
note that recent results \cite{guimaraes} for 2+1
gravity coupled with non-Abelian Chern-Simons vortices on an axially-symmetric
spacetime indicate Bogomolny equations are possible (subject to appropriate
ansatz) for particular (and in this case nonrenormalizable) potentials.
It is an interesting question to see whether the methods of this
paper will allow an extension to nontrivial potentials.

Having obtained the Bogomolny equations we have also considered their
solutions. Two cases arise depending on whether the $\sigma$-model field
is covariantly constant or not.  The former situation is somewhat
easier to analyse and leads to a set of (Euclidean)
Abelian magnetostatic equations,
valid for any gauge group. These equations are sufficient
to determine solutions of the Bogomolny equations in this case.

When the $\sigma$-model field is not covariantly constant the Bogomolny 
equations are rather more complicated. Here we have focussed on  
axially symmetric solutions for the
case of $su(2)$, though the  situation involving larger gauge groups is
very interesting. 
By projecting onto the various gauge components
we  may express the Bogomolny as the coupled system of first-order
equations (\ref{Bogform}). 
The imposition of axial symmetry on space places additional constraints
on the  parameters of an axially symmetric gauge connection.
At this stage we adopted an ansatz\footnote{One hope is to find an 
integrable system in these equations paralleling the appearance of the 
Toda equations in the flat space monopole equations.} 
to simplify the analysis of (\ref{Bogform}) and we recovered the
results of previous authors who studied the spherically
symmetric case.
In fact we were able to show that an asymptotically flat, spherically
symmetric black hole solution with regular gauge fields is necessarily
an extreme Reissner-Nordstr\"{o}m black hole.
Our work also shows a totally different type of radial dependence between
the higher multipoles and the  monopole terms in our axially symmetric solution.
Interestingly an improved asymptotic approximation
suggested a \lq massive magnetic core\rq, the asymptotic corrections having
the same form as instanton corrections.
In our final section we extended the original ansatz to include higher
winding numbers finding an exact, covariantly constant solution for 
arbitrary winding number. 
We have not succeeded thus far in finding solutions to this ansatz for
non-covariantly constant solutions with higher winding numbers.
Although obtaining exact, nonlinear solutions may be an 
impossible task, expansions
in powers of $\frac{1}{r}$ and $\theta$ (in the spirit of \cite{kleihaus})
could help to understand some aspects of the
solutions. Such an approach is left for future work.

Clearly obtaining the Bogomolny equations is far from the
end of the matter. Although this system of equations is  of lower
order and simpler than the original field equations, they are still
complicated. Our analysis has presented some of the features of these
equations and we look forward to their wider study.

\section*{Acknowledgements}
HWB is grateful to E.J. Weinberg and D. Maison for numerous discussions.
Victor Varela is grateful to the
Mathematics and Statistics Department of Edinburgh University
for its hospitality during his long-term research visit.
He also wishes to thank those attending the Applied Mathematics
Seminar for their comments on an earlier version of this paper
and acknowledges the use of the MAPLE program at various stages in this
work.
We thank the referee for suggested improvements to the paper.

\section*{Appendix: Remarks on Singularities}
In this appendix we consider the possibility of finding non-singular solutions 
to the gravitationally coupled Bogomolny equations, restricting our attention
to the case of vanishing dilaton  ($\gamma=0$).

Besides the invariants $R$, $R_{\alpha\beta}R^{\alpha\beta}$, 
and $R_{\alpha\beta\mu\nu}R^{\alpha\beta\mu\nu}$, the simplest one that can be
used in the analysis of singularities is $J=F^{a}_{\mu\nu}F^{a\mu\nu}$, which
is the non-Abelian generalisation of the one used by Hartle and Hawking in 
their study of Abelian, multi-black hole solutions\cite{HH}.
Using (\ref{Bogomolny1}) we obtain the expression
\begin{equation}
J=2\left(D_{i}n^{a}D^{i}n^{a}+\partial_{i}u \partial^{i}u \right).
\label{scalar}
\end{equation}
If we assume that $n$ is covariantly constant and choose the metric 
(\ref{isometric}),
then (\ref{scalar}) reduces to $J=2\left(\nabla V \right)^2$. In this special
case $1/V$ is harmonic (see Section 3) and can be determined 
exactly even without
spatial symmetries. One solution is
\begin{equation}
\frac{1}{V}= 1+\sum_{p=1}\sp{N}\frac{|m_{p}|}{\rho_{p}},
\label{multibh}
\end{equation}
where $\rho_{p}=\sqrt{\left(x-x_{p}\right)^{2}+\left(y-y_{p}\right)^{2}
+\left(z-z_{p}\right)^{2}}$.
This solution describes a collection of $N$ point-like monopoles with
magnetic charge $m_{p}$, located at $\left(x_{p},y_{p},z_{p}\right)$.
For stationarity every $m_{p}$ has the same sign and (\ref{multibh})
describes the
Majumdar-Papapetrou solutions considered in \cite{HH} and $J$ is finite 
everywhere except at some points which are enclosed by event
horizons. Also, according to Hartle and Hawking, other distributions of charge
different to discrete point sources can be considered. 
However, the corresponding solutions can be shown to have naked singularities.
We conclude that the covariantly constant solutions of (\ref{Bogomolny1}) 
are necessarily singular,
but these (genuine) singularities are not naked when the source is a 
discrete set of (like charged) point-like magnetic charges as described by
(\ref{multibh}).

The analysis of singularities in the non-covariantly constant case is much
more difficult because $1/V$ is not harmonic and we do not know any exact,
non-trivial solution of the non-linear equations (\ref{Bogomolny1}). 
Nevertheless, we present some considerations based on our knowledge of 
asymptotic solutions. 

Restricting our attention to the axially symmetric case, we use spherical
coordinates and (\ref{abc}) to  obtain the reduced expression
\begin{equation}
J=\frac{4 V^{2} \chi^{2}}{r^2}+2\left(\nabla V \right)^2.
\label{nonabj}
\end{equation}
Both terms in (\ref{nonabj}) are non-negative, so there is no
possible cancellation of infinities and, exactly as in the covariantly
constant case, the solution is singular wherever $1/V=0$.

Using (\ref{bogop1}), (\ref{solbawa}) and (\ref{nonabj}) it is easy to 
see that $J$ is unbounded 
at each one of the infinite critical points of $\ln{|\chi(r)|}$ near $r=0$
thus yielding genuine singularities.  One may ask whether there exist
spherically symmetric solutions of (\ref{pdechi}) 
with such pathological behaviour for very small $r$
yet that behave well in the far asymptotic region.
We have numerical evidence for the existence of such a class of
spherically symmetric, asymptotically flat solutions.
The transition between the (asymptotic) monotone and
oscillatory regimes of (\ref{appse}) and
(\ref{solbawa}) can be conveniently
analysed in terms of the new variable $\xi( x)=\ln \chi^{2}$,
where $ x=\ln r$, which satisfies the equation
\begin{equation}
\frac{\partial^{2}\xi}{\partial x^{2}}-
\frac{\partial\xi}{\partial x}+2e^{\xi}=2.
\label{autode}
\end{equation}
This transformation enables a simple numerical analysis
near $ x=0$. The solution shown in Figure 1, satisfies
$\xi (0)=1,\:\xi^{\prime}(0)=0$. It evidences the different behaviours
of (\ref{appse}) and (\ref{solbawa}),  and displays a smooth regime transition
at $ x=0\:(r=1)$. As a consequence of (\ref{bogop1}), $V^{2}$ is
infinite at every $ x$ where $\xi^{\prime}( x)=0$.
It appears then that (\ref{pdechi}) admits at least one class of 
asymptotically flat, non-linear solutions
which contain essential singularities.

We next would like to extend our analysis of singularities to 
non-linear solutions in the non-spherically symmetric case.
This task is even more difficult because at present we
only know the asymptotic behaviour of putative solutions and have not
answered the question of existence.
Equations (\ref{asypaxi})-(\ref{axipsi}) show the 
compatibility of (\ref{pdechi}) with an asymptotically flat spacetime, and 
(\ref{axichi}) exhibits
the asymptotic behaviour of any possible solution satisfying 
$|\chi(r,\theta)|\approx 1$ near $r=0$.
As in the spherically symmetric case, the (oscillatory) monopole term in 
$\ln{|\chi(r,\theta)|}$ leads to spacetime singularities which can be avoided
only if $C=0$ in (\ref{axichi}). Also, it is easy to see that the new leading 
$l=1$ term
makes $V$ behave as $\frac{1}{|cos\theta|}$ when $r\rightarrow 0$.
This asymptotic behaviour also implies unbounded values for $J$ near $r=0$ as a 
consequence of (\ref{nonabj}), so we also put $c_{1}=0$ in (\ref{axichi}) 
and proceed with the 
analysis of the remaining terms in that series. The
 asymptotic behaviour of the higher multipole terms $\left(l=2,3,4,...\right)$
 once more implies that $1/V$ vanishes at $r=0$,  and so  at that
point $J$ is infinite again.
Clearly, this implies the existence of spacetime singularities 
arising from solutions which have the asymptotic 
behaviour $|\chi(r,\theta)|\approx 1$ near $r=0$.
However, we can say nothing about the existence of such singularities 
in solutions with a different type of asymptotic behaviour. 
We suspect that every asymptotically flat solution must give
rise to spacetime singularities near $r=0$.
We doubt that analytic methods will
be useful at this stage for studying the solution space of (\ref{pdechi}), so
the next step should be a detailed numerical study of possible
non-spherically symmetric solutions.

\newpage
 
\begin{figure}[h]
 
\begin{center}
\scalebox{0.5}[0.5]{\includegraphics*[bb=88 105 525 701]{fig1.eps}}
\caption{\label{fig1}Numerical solution for equation (\ref{autode}).}
\end{center}
 
\end{figure}

\end{document}